\documentclass[aps,twocolumn,prd,showpacs,nofootinbib]{revtex4}
\usepackage{amsmath}
\usepackage{graphicx}
\usepackage{dcolumn}
\usepackage{bm}
\usepackage{amssymb}
\usepackage{latexsym}

\def\be{\begin{equation}}
\def\ee{\end{equation}}
\def\ba{\begin{eqnarray}}
\def\ea{\end{eqnarray}}

\begin{document}

\title{Tunnelling for Large N}

\author{Yun-Song Piao}

\affiliation{College of Physical Sciences, Graduate University of
Chinese Academy of Sciences, Beijing 100049, China}

\begin{abstract}

In this brief note, by applying the stochastic approach to multiple
fields, we estimate the probability of tunneling between vacua in
a landscape with $\cal N$ fields. We find that the probability can
be enhanced by large $\cal N$. When $\cal N$ saturates the dS
entropy of some vacuum, the corresponding vacuum will be extremely
unstable and be expected to rapidly decay. We discuss the
implications of this result.

\end{abstract}

\maketitle

Recently, the string landscape with large number of vacua has
received increased attentions \cite{S}.
The cosmological dynamics in the landscape is exhibited as eternal
inflation \cite{vilenkin, linde}, see Refs.\cite{V06, S06} for reviews, during which multiple universes
with various vacua were spawned. Thus it seems that the thing left
is how to make the predictions in such a picture and compare them
with observations. However, before proceeding, it might be
significant to recheck whether the occurrence of tunneling is
actually as imagined as suppressed exponentially \cite{GW, S1983},
since the tunneling histories can affect the measure for making
the predictions \cite{AGJ}. In the low energy, the landscape can be
visualised as a multiple dimensions or fields space with a
complicated and rugged potential. Thus the tunneling in the
landscape will inevitably involve the fluctuations of many fields,
or moduli. This means it is necessary to include the evolutions or
random jumps of lots of fields in a referred tunneling event.

The inclusion of multiple fields will possibly inflect the result
for tunneling. For example, in Ref. \cite{JL} it was showed that
some of bubble solutions contain naked timelike singularities,
which will render the corresponding vacuum more stable. However,
in this brief note we will study this issue by applying the
stochastic approach for multiple fields. In the stochastic
approach \cite{AAS, GLM}, the field can climb up to the top of
potential barrier by the random walk induced by quantum
fluctuation during inflation, and then roll down along other sides
to new vacua, which in some sense corresponds to a tunneling. The
probability finding the field at some position in the stochastic
approach rests with its drift parameter and diffusion parameter.
The diffusion parameter is determined by the quantum fluctuations
of fields, and denotes the step length of random walk in unit of
Hubble time. In principle, the larger it is, the higher the
probability is. In the case of multiple fields, the quantum effect
will certainly enhanced. This enhancement has lead to some
interesting effects. It has been noticed \cite{APQ2, APQ3} that
there is a large N phase transition near some critical value,
beyond which the slow roll inflation phase will disappear. This
result is also consistent with recent arguments from black hole
physics \cite{D07}, and also in \cite{Huang, LS1} for relevant
studies for multiple field inflation. Thus it can be reasonably
expected that large N might drastically alter our understanding to
the tunneling.

\begin{figure}[t]
\begin{center}
\includegraphics[width=6.5cm]{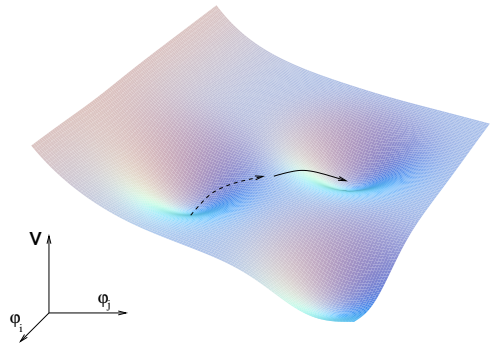}
\caption{The landscape of a potential with multiple fields, i.e.
dimensions. The potential is the sum of all fields contributing
the vacua, in which for each field their potentials do not must be
same. The tunneling described by the stochastic approach
corresponds that the fields stochastically climb up to the top of
potential barrier, which is induced by the quantum fluctuations of
fields, see the dashed line, and then roll down along other sides
to new vacua, see the solid line.}
\end{center}
\end{figure}

We begin with the multiple dimensions stochastic equation. The
probability finding the fields in their corresponding positions at
some time is \cite{FP} \be {\partial{\cal P}\over
\partial t}= \sum_{i,j}{\partial^2 ({\cal D}^{(2)}_{ij}{\cal P})\over
\partial \varphi_i\partial \varphi_j}-\sum_i{\partial ({\cal D}^{(1)}_i{\cal P})\over \partial
\varphi_i}, \label{allpequ}\ee where ${\cal D}^{(1)}_i$ is the
drift vector and ${\cal D}^{(2)}_{ij}$ is the diffusion tensor,
and both are the functions dependent of all $\varphi_i$. The
referred $\varphi_i$ denotes the field being concerned with
building the corresponding vacua. Thus the potential in field
space is the sum of potentials of all fields $\varphi_i$ together
building the vacua. In principle the potential of each field do
not must be same, since we only require that all them have the
contributions to build vacua and do not care how much the
contribution of each field is. The Fig.1 is a sketch for an
illustration. Eq.(\ref{allpequ}) is actually difficult to be
solved. Thus we will consider the radial motion of fields for an
estimate to the probability of tunneling, and expect that such a simplified operation can capture
the general characters of full results. The potential is assumed
to be broad so that the slow roll condition can be applied and the
random walk is feasible. The square of $ d\varphi$ in fields space
is the square sum of all $d\varphi_i$, in which $\varphi$ denotes
the equivalent radical field. The slow roll equation for each
field is $3h{\dot \varphi_i}+V_i^{\prime}\simeq 0$, where $h$ is
the Hubble parameter. Thus after multiplying $d\varphi_i$ to this
equation and then making the sum for equations of all fields, we
have $3h{\dot \varphi}+ { V}^{\prime}\simeq 0$, in which $V$ is the potential in
field space and corresponds to the sum of potential $V_i$ of all fields.
This means the
radial motion in field space can be described equally by an
effective radial field obeying the same slow roll equation with
that of each field. Thus the motion of this radial field can be
given by one dimension stochastic equation, which corresponds to
that of Eq.(\ref{allpequ}) with only one freedom and equals to
that used in Refs.\cite{AAS, GLM}.


When there is only one freedom, ${\cal D}^{(1)}$ and ${\cal
D}^{(2)}$ will be not vector and tensor any more, respectively. In
this case. ${\cal D}^{(1)}(\varphi)$ may be called the drift
coefficient of radial motion, which here equals to $\dot \varphi$,
and ${\cal D}^{(2)}(\varphi)$ is the diffusion coefficient of
radial motion, which can be determined as follows. In slow roll
approximation, each field contributes the fluctuation $\delta
\varphi_i\simeq {h\over 2\pi}$, thus the radial random walk in field
space has the step length
$(\delta\varphi)^2 \simeq {\cal N}({h\over 2\pi})^2$ for each time
interval $\Delta t\sim 1/h$. In this case, similar to the
calculations in Ref. \cite{LLM}, \be {\cal D}^{(2)}(\varphi)\simeq
{\cal N}{h^3({\varphi})\over 8\pi^2} \label{cald}\ee can be found.
It can be noticed that compared with single field, here the
diffusion coefficient ${\cal D}^{(2)}$ is markedly enhanced by the
number $\cal N$ of fields, which means that compared with single
field, the step length of effective random
walk in field space is amplified by $\cal N$.

The stationary solution can be found by taking ${\partial {\cal
P}\over \partial t}=0$. Thus the probability finding $\varphi$ in
some position is \be {\cal P}\sim \exp{\int{{\cal D}^{(1)}\over
{\cal D}^{(2)}}d\varphi}\simeq\exp{ \int {8\pi^2{\dot
\varphi}\over {\cal N}h^3(\varphi)}d\varphi}, \label{pree}\ee
where Eq.(\ref{cald}) has been applied. The probability of
$\varphi$ jumping the top of its potential barrier can be obtained
by making the integral for $\varphi$ from its original background
vacuum to the top of its potential barrier, which is \be {\cal P}
\sim
\exp{ \left(-{3\over 8{\cal N}}({1\over { V}_{\rm b}} -{1\over
{V}_*})\right)}, \label{pre}\ee where the gravitational scale $G=1$ has
been set, and the subscript `b' and `$\ast$' denote the values of
original background vacuum and the top of its corresponding
potential barrier, respectively. For ${\cal
N}=1$, Eq.(\ref{pre}) is exactly same with that for single field
\cite{GLM, LLM}, which is suppressed exponentially. While for
large $\cal N$ the result given by Eq.(\ref{pre}) is
enhanced. This means that the probability tunneling to adjacent
vacua is actually heightened by large $\cal N$. This result is not surprised, since
it is straightly leaded by Eq.(\ref{cald}), in which the diffusion
coefficient has a large $\cal N$ enhancement. In contrast, if the
diffusion coefficient has a suppression, the probability of
tunneling will be depressed, see Ref. \cite{CW} for studying in
noncommutative eternal inflation.

When ${\cal N}\simeq {1\over V_{\rm b}}$, the probability ${\cal
P}\sim 1$, since generally $V_{\rm b}\ll V_*$. Further, noticing
the dS entropy of background vacuum $S\simeq {1\over h^2}\simeq
{1\over V_{\rm b}}$, thus we obtain that when ${\cal N}\simeq S$,
${\cal P}\sim 1$. This means when the number of fields saturates
the dS entropy of vacuum which all these fields builded together,
the corresponding vacuum will be overly unstable, and will rapidly
tunnel to other neighboring vacua by quantum jump. This result is
consistent with that obtained in Ref. \cite{APQ2, APQ3} for
Nflation. In Refs. \cite{APQ2, APQ3}, it is found that there is a
critical point for large N transition, in which the number of
fields saturates the dS entropy, beyond which the quantum effect
will strong so that the slow roll inflation phase disappears. Here
the enhancement of quantum effect is reflected in the increase of
the diffusion coefficient ${\cal D}^{(2)}$. It seems that when
${\cal N}\simeq S$, ${\cal D}^{(2)}$ completely overwhelms the
drift coefficient ${\cal D}^{(1)}$, and the latter is
negligible. In this case, it is interesting to notice that in unit
of Hubble time the diffusion distance in field space $\sqrt{{\cal
D}^{(2)}\Delta t} \sim 1$, where ${\cal N}\simeq S\sim 1/h^2$ and
$\Delta t\simeq 1/h$ have been applied. This means this diffusion
distance in a landscape of field space is in Plank order, which is enough large for a jump to
the top of potential barrier and thus the tunneling. In another point of view, it can be
intuitively thought that there is at least a freedom degree for
every field, thus the total freedom degree of $\cal N$ fields
system, i.e. the entropy, should be at least $\cal N$. Thus a
vacuum state with ${\cal N}>S$ will be obviously impossible, since
$S$ is the dS entropy which is the maximal entropy of
corresponding system. This reflect the fact again why such vacua
should rapidly decay.

For Eq.(\ref{pre}), if ${\cal N}> {1\over V_{\rm b}}$, then the
corresponding vacuum will be unstable and rapidly tunnel to other
vacua. Thus in this sense, the number $\cal N$ of fields can be
thought as a division to violently unstable and metastable vacua.
For fixed $\cal N$, only the vacua with the energy density small
than $1/{\cal N}$ are metastable, while those large than $1/{\cal
N}$ are violently unstable, which will experience a or a series
rapid tunneling till a metastable vacuum is arrived. The value
$1/{\cal N}$ corresponds to set a critical energy density, beyond
which the vacuum will be not possible to be stable or metastable.
For Eq.(\ref{pre}), the critical energy density for single field,
i.e. one dimension landscape, is $1/{\cal N}\sim 1$, which is
Planck order. This in some sense is also the reason why the tunneling is exponentially
suppressed. In order to make a critical energy density slightly
larger than the cosmological constant value observed currently,
${\cal N}\sim 10^{120}$ is required, which seems not realistic.
However, it is plain that a lower critical energy density can be
acquire by considering a larger $\cal N$. When the energy density
of successive vacua satisfies $V_{\rm b}\gtrsim 1/{\cal N}$, the
tunnelings will be expected to occur one after the other with the
probability ${\cal P}\sim 1$, up to the vacuum with $V_{\rm b}<1/{\cal
N}$ in which the rapid tunneling ends. In this case, the scenario
will be slightly similar to that of chain inflation \cite{FS,
Huangchain, CD}. However, here the tunneling is regarded in a
different viewpoint. Thus the relevant scenario will be different, and
need to be
explored.

The inclusion of multiple fields, in some sense,
corresponds to reduce the effective gravitational scale in
corresponding model with single field
\cite{D07}. Thus the stochastic approach and the method of HM instanton \cite{HM} seems
possibly give same results for large $\cal N$, which is valid for single field \cite{GLM, LLM}.
However, recently, it was showed in \cite{JL2, JL} that with
multiple fields, in many classes of potentials there dose not
exist the HM points, since it is not easy to find a top of
barrier which is an extremal point for all fields. It is generally thought that
the HM instanton occur when
the gravitational effect is dominated. Thus it seems that there
maybe not the tunneling in such a strong gravitational regime.
This seem conflicted with the result obtained here, in which the
tunneling will not only occur but be enhanced, especially the
stronger the gravitational effect is, which is charactered by the
large number of fields, the easier the tunneling is.
The underlying
reason of this conflict might be that the application of the stochastic
approach only requires there are the barriers separating the
vacua, and is not dependent of whether there is a top of barrier
behaving the extrema for all fields, as the HM instanton requires.

It can be also noticed that, even if there is a top of barrier
behaving the extrema for all fields, in the stochastic approach the random walk of fields
also must not pass through this top, since in
multiple dimension field space the paths for fields striding over
the barrier between both vacua is certainly not only one. The different
paths, i.e. different top ${1\over V_*}$, will give different
probability, though ${1\over V_{\rm b}}$ is actually dominated. In
principle, the fields will most possibly pass though the top
giving the largest probability, which must not coincide with that
behaving the extrema for all fields. Thus in these cases whether the results from
the stochastic approach and the instanton method is consistent
seems still open. These results and observations indicate that
in a field space with large $\cal N$
fields, the tunneling between vacua seems more subtle than imagined.

In summary, by applying the stochastic approach to the
landscape with $\cal N$ fields, we found that the tunneling
probability between vacua can be enhanced by large $\cal N$.
The more the number of fields is, the
larger the probability is. When $\cal N$ saturates the dS entropy
of some vacuum, the corresponding vacuum will has the decay
probability ${\cal P}\sim 1$, and thus will be extremely unstable
and be expected to rapidly decay. The implications of this result were discussed.
It should be mentioned that in
order to have an estimate for the probability we identify the problem be
effectively one dimensional. Thus it is inevitable that the result obtained is
slightly rough. 
which, however, might
have captured some essentials of full answer.
We hope this work could be an interesting step towards a
comprehensive understanding to the tunneling in the landscape with
large $\cal N$ dimensions. There are many open issues left to
study.

\textbf{Acknowledgments} The author thank Y.F. Cai for helpful
discussions and comment. This work is supported in part by NSFC
under Grant No: 10405029, 10775180, in part by the Scientific
Research Fund of GUCAS(NO.055101BM03), in part by CAS under Grant
No: KJCX3-SYW-N2.

\end{document}